\documentclass[aps,twocolumn,preprintnumbers,letter]{revtex4}
\usepackage[dvips]{graphicx}

\begin{document}
\title{Thermalization in SU(3) gauge theory after a deconfining quench}

\author{Alexei Bazavov}
\affiliation{Department of Physics, 
University of Arizona, Tucson, AZ 85721, USA}
\author{Bernd A. Berg}
\affiliation{Department of Physics, 
Florida State University, Tallahassee, FL 32306, USA}
\author{Adrian Dumitru}
\affiliation{Institut f\"ur Theoretische Physik,
Johann Wolfgang Goethe Universit\"at,
Max-von-Laue-Str.\ 1,
D-60438 Frankfurt am Main, Germany}

\begin{abstract}
We determine the time evolution of fluctuations of the Polyakov loop
after a quench into the deconfined phase of SU(3) gauge theory from a
simple classical relativistic Lagrangian. We compare the structure
factors, which indicate spinodal decomposition followed by relaxation,
to those obtained via Markov Chain Monte Carlo techniques in SU(3)
lattice gauge theory. We find that the time when the structure factor
peaks diverges like $\sim 1/k^2$ in the long-wavelength limit. This is
due to formation of competing Z(3) domains for configurations where
the Polyakov loop exhibits non-perturbatively large variations in
space, which delay thermalization of long wavelength modes. For
realistic temperatures, and away from the extreme weak-coupling limit,
we find that even modes with $k$ on the order of $T$ experience
delayed thermalization. Relaxation times of very long wavelength modes
are found to be on the order of the size of the system; thus, the
dynamics of competing domains should accompany the hydrodynamic
description of the deconfined vacuum.
\end{abstract}

\maketitle

\section{Introduction}

Relativistic Heavy-Ion Collision (RHIC) experiments carried out at
Brook\-haven National Laboratory (BNL) provide support for the
existence of a quark-gluon plasma (QGP) phase of QCD~\cite{QGP}. In
this phase color charges are liberated (deconfined), contrary to the
low temperature phase that confines color charges inside colorless
objects such as hadrons or glueballs. The existence of confined and
deconfined phases has been demonstrated
numerically in Lattice Gauge Theory (LGT) studies \cite{Tc81}. Lattice
QCD predicts a first-order phase transition for SU(3) pure gauge
theory \cite{SU3Tc} that turns into a crossover for full
QCD~\cite{QCDTc} with physical quark masses.

A collision of two heavy nuclei at high energy releases a large number
of gluons from the wave functions of the colliding
nuclei~\cite{MG_McL}. Those gluons interact and eventually form a
thermalized QCD plasma with a temperature in excess of the critical
temperature for deconfinement. Complete (and consistent) theoretical
understanding of the thermalization process is presently lacking.
Baier, Mueller, Schiff and Son developed the so-called ``bottom-up''
approach~\cite{BottomUp}, which is a framework for understanding the
processes leading to thermalization and for calculating the
thermalization time of the QCD medium as well as its initial
temperature (see, however, the critique in ref.~\cite{ALM04}).  
The ``bottom-up'' approach is based on solving
the Boltzmann equation for quasi-particles in a trivial vacuum and
neglects the structure of the deconfined phase of the non-Abelian
gauge theory arising from the Z(3) center symmetry discovered by
t'~Hooft and Polyakov~\cite{tHooft}. Below, we shall show that in a
model which allows for non-perturbatively large variations of Polyakov
loops in space, domain walls form which separate regions of
different Z(3) orientation~\cite{Z3domains}. The competition between
such domains affects the thermalization of long-wavelength modes of
the Polyakov loop. We shall also provide some model estimates for the
relevant wave lengths and time scales.

We adopt a simplified picture where a relativistic heavy-ion
collision is viewed as a quench that instantly heats the system
to a temperature above the deconfining temperature. The response of
Polyakov loop Structure Factors (SFs) to such a heating quench has been
studied in SU(3) LGT by Markov Chain Monte Carlo (MCMC)
simulations for Glauber (dissipative) dynamics~\cite{BBV06}. As
unambiguous signals for the transition one finds a dynamical growth of
the SF, reaching maxima which scale approximately with the volume, a
behavior often characterized as spinodal decomposition~\cite{MO00}.

Glauber (model~A) dynamics~\cite{Gl63} imitates the thermal
fluctuation of Nature. It is therefore expected to describe the
dissipative features of the transition from one equilibrium ensemble
to another well. MCMC updating with a Metropolis or heatbath algorithm
falls into the universality class of model~A. Such SU(3) MCMC
simulations converge to 3D equilibrium ensembles, which are the same
as in Minkowski space, because the fourth extension of the Euclidean
lattice serves only to define the temperature. The major drawback of
Glauber dynamics is not the 4D Euclidean formulation but that it is
non-relativistic and, more importantly, that one does not know how to
connect the MCMC updating step to a physical time scale.

It was stressed in \cite{BBV06a} that relaxation to the vacuum
ensemble at high temperature becomes feasible only after each SF has
overcome its maximum value. For SU(3) gauge theory this relaxation
time diverges with increasing system size due to competing order-order
domains with different Z(3) center group triality, which are similar 
to order-order domains in a 3-state Potts model~\cite{SY82}. This 
divergence is well known in condensed matter
physics~\cite{CL97}. Hence, one should not a-priori exclude the
possibility that, under heating, the long wavelength modes in the
system do not equilibrate but instead get stuck in the neighborhood of
the SF maxima.

In this paper we explore heating quenches into the deconfined phase
within Pisarski's effective model of Polyakov loops \cite{Pi00} (see
also ref.~\cite{MFK07} for similar effective potential models), which
we compare to those from Glauber dynamics. The solutions of the 
effective theory are obtained through Molecular Dynamics (MD) 
simulations of hyperbolic, relativistic Minkowski dynamics \cite{Du01} 
for which the time scale of the updating step is known (simulations 
based on a Langevin approach with statistical noise and friction have 
also been performed, see ref.~\cite{Langevin}). 

In Sec.~\ref{sec_eff_model} we summarize the model and explain the 
simulations.  Sec.~\ref{sec_num_res} presents our numerical results.  
Conclusions and outlook follow in Sec.~\ref{sec_concl}.


\section{Effective Model of Polyakov Loops}\label{sec_eff_model}

In Pisarski's model \cite{Pi00} the deconfined phase of a pure gauge
theory is described as a condensate of Polyakov loops. The
Z(3)-symmetric effective potential for Polyakov loops $\ell$ (complex
for SU(3)) with cubic and quartic interactions takes the form
\begin{equation} \label{ellpot}
  {{{\cal V}(\ell)=\left( -\frac{b_2}{2}
  |\ell|^2-\frac{b_3}{6}(\ell^3+(\ell^*)^3)+\frac{1}{4}
  (|\ell|^2)^2\right) b_4\,T^4}}\ .
\end{equation}
The energy scale is set by $T^4$, and the mass coefficient $b_2=b_2(T)$ 
is temperature dependent, while $b_3$ and $b_4$ are constants. These 
couplings can be chosen so that they reproduce lattice data for the SU(3) 
pressure and energy density above $T_c$. A reasonable fit follows with 
$b_4=0.61\,r^4$, $b_3=2.0/r$, and $b_2(T)=((1-1.11/x)(1+0.265/x)^2
(1+0.300/x)^3-0.487)/r^2$, where $x\equiv T/T_c$ and $r=2.23$.
A plot is shown in fig.~1 of ref.~\cite{Kpi}.

To complete the effective theory in Minkowski space-time, a kinetic term
has to be added:
\begin{equation}\label{ec}
  {\cal L} \; = T^2 \left( Z_t \, |\partial_t \ell|^2 + Z_s \,
    |\partial_i \ell|^2 \right) - {\cal V}(\ell)\ .
\end{equation}
Here, we assume a Lorentz-invariant form, $Z_t=Z_s$, and take the
coefficient $Z_s$ from that for spatial variations of SU(3) Wilson
lines~\cite{Pi00}, $Z_s=N_c/g^2$ with $g^2=3$. Thus the dynamics of
Polyakov loops is in an intermediate regime between very weak
($Z_s\gg1$) and very strong ($Z_s\ll1$) coupling.

\begin{figure}[h] \begin{center}
\includegraphics[width=\columnwidth]{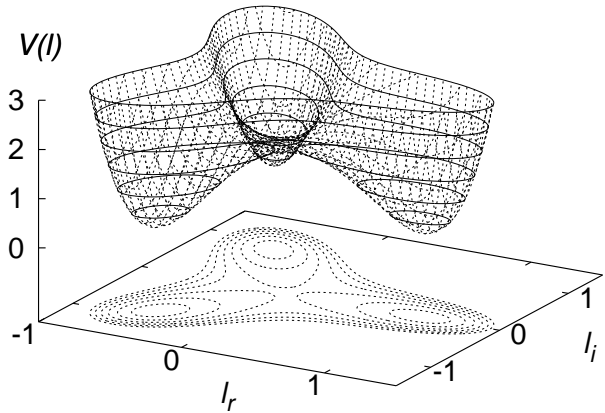}
\caption{Effective potential for the Polyakov loop $\ell$ from
  eq.~(\ref{ellpot}) at $T_f/T_c=2.0$, shifted by a constant.
Contours of equal heights are shown in the $(\ell_r,\ell_i)$ plane.} 
\label{fig_ploop_pot} 
\end{center} \end{figure}

We employ a simulation procedure similar to the one of ref.~\cite{Du01} 
but focus on heating quenches of the system and not on its subsequent 
cooling \cite{TV05}; also, we consider here a static, non-expanding 
metric. Polyakov loop fields are defined on the sites of
a spatial cubic lattice of size $N_s^3$ with periodic boundary
conditions. They are initialized in the confined phase at time
$t=0$. Then the temperature entering the effective
Lagrangian~(\ref{ec}) is set to a value $T_f>T_c$ above the
deconfinement transition $T_c$, where the Z(3) center symmetry is
spontaneously broken. At temperatures $T_f>T_c$ the effective
potential takes the shape shown in Fig.~\ref{fig_ploop_pot}. In the
plane below we show equal height contours.

After the quench to $T_f$ the system evolves in time according to the 
Euler-Lagrange equations derived from the Lagrangian~(\ref{ec}) and 
rearranges itself to a new equilibrium ensemble with non-zero Polyakov 
loop, which signals symmetry breaking in the infinite volume limit.

In the following we describe how the initial field configurations are 
constructed.

\subsection{Initialization of the field}

Introducing real and imaginary parts, we write the Polyakov loop as
\begin{equation}\label{lri}
    \ell=\ell_r+i\,\ell_i~.
\end{equation}
We split the fields into a long- and a short wavelength part,
\begin{equation}\label{Llongshort}
  \ell_r = \bar\ell_r + \delta\ell_r~~,~~ 
  \ell_i = \bar\ell_i + \delta\ell_i~.
\end{equation}
The system is initialized in the unstable confined phase, thus 
\begin{equation}
  \bar \ell_r(\vec{x},t=0)=\bar \ell_i(\vec{x},t=0)=0\ .
\end{equation}
The initial fluctuations are assumed to be Gaussian,
\begin{equation}\label{PlrPli}
  P[\ell_r] \sim \exp
  \left( - \frac{\ell_r^2}{2\sigma_r^2}\right)~~,~~
  P[\ell_i] \sim \exp
  \left( - \frac{\ell_i^2}{2\sigma_i^2} \right)~.
\end{equation}
Consequently, $\langle \delta\ell_r \rangle=\langle \delta\ell_i
\rangle=\langle \delta\ell_r \delta\ell_i\rangle=0$,
$\langle\delta\ell_r^2 \rangle=\sigma_r^2$ and 
$\langle\delta\ell_i^2 \rangle=\sigma_i^2$ hold.

\begin{widetext}
\subsection{Counterterms in the equations of motion}\label{sec_EoM}

In terms of $\ell_r$ and $\ell_i$, the potential can be
written as:
\begin{equation} 
{\cal V} 
\, =\, b_4 T^4 \left[ -\frac{b_2}{2}(\ell_r^2+\ell_i^2) -\frac{b_3}{3}
  \ell_r(\ell_r^2 -3\ell_i^2)
  + \frac{1}{4}(\ell_r^4+\ell_i^4+2
  \ell_r^2 \ell_i^2) \right]~. \label{PotReIm}
\end{equation}
Using eq.~(\ref{Llongshort}) and averaging over the distribution of
initial fluctuations (\ref{PlrPli}) up to order $\delta\ell^2$ gives
the one-loop correction to the effective potential: 
\begin{eqnarray}
 {\cal V} &
= & b_4 T^4 \left[ -\frac{b_2}{2}(\bar{\ell_r}^{2}+\bar{\ell_i}^2) -
\frac{b_3}{3} \bar{\ell_r}(\bar{\ell_r}^2 -3\bar{\ell_i}^2) +
\frac{1}{4}(\bar{\ell_r}^4+\bar{\ell_i}^4+2\bar{\ell_r}^2
\bar{\ell_i}^2) \right] \nonumber\\
 &+& b_4T^4\left[
b_3\bar{\ell_r}(\langle \delta\ell_i^2\rangle - \langle
\delta\ell_r^2\rangle) + \frac{3}{2} (\bar{\ell_r}^2\langle
\delta\ell_r^2\rangle + \bar{\ell_i}^2\langle \delta\ell_i^2\rangle) +
\frac{1}{2} (\bar{\ell_r}^2\langle \delta\ell_i^2\rangle +
\bar{\ell_i}^2\langle \delta\ell_r^2\rangle)\right] + {\rm const}\ .  
\end{eqnarray}
The additional second term, which arises from the classical initial
fluctuations, needs to be subtracted in order to restore the original
potential~(\ref{PotReIm}) for the long wavelength modes of the
Polyakov loop. The equations of motion including these counterterms
are given by
\begin{eqnarray} \nonumber
Z_s T^2  \partial_\mu\partial^\mu\ell_r+\frac{1}{2}\frac{\partial{
  \cal V}}{\partial\ell_r} - \frac{b_4T^4}{2}\left[
  b_3(\langle \delta\ell_i^2\rangle-\langle \delta\ell_r^2\rangle)
  + 3{\ell}_r\langle \delta\ell_r^2\rangle
  + {\ell}_r\langle \delta\ell_i^2\rangle\right] &=& 0\,,\\
Z_s T^2   \partial_\mu\partial^\mu\ell_i+\frac{1}{2}
    \frac{\partial{\cal V}}{\partial\ell_i} - 
    \frac{b_4T^4}{2}\left[
    3{\ell}_i\langle \delta\ell_i^2\rangle
    +{\ell}_i\langle \delta\ell_r^2\rangle\right] &=& 0\,, 
\end{eqnarray}
where we replaced $\bar \ell$ with $\ell$ because the difference in the 
equations of motion is of order $\delta \ell^3$ and can be neglected.
In the simulation we take $\sigma_r^2=\sigma_i^2=\sigma^2$ so that the 
counterterms simplify to 
\begin{eqnarray}\label{eom_lr}
Z_s T^2   \partial_\mu\partial^\mu\ell_r + \frac{1}{2}\frac{\partial{\cal
   V}}{\partial\ell_r} - 2b_4T^4 {\ell}_r \sigma^2 &=& 0~,\\
\label{eom_li}
Z_s T^2   \partial_\mu\partial^\mu\ell_i + \frac{1}{2}\frac{\partial{\cal
  V}}{\partial\ell_i} - 2b_4T^4 {\ell}_i \sigma^2 &=& 0~.
\end{eqnarray}
\end{widetext}
The numerical results shown below were obtained with $\sigma=0.04$; 
$\sigma=0.08$ gives similar results (our statistical errors are larger
than the systematic errors due to variation of the initial
fluctuations within this range). We do not have a good 
quantitative estimate for the magnitude of fluctuations, but it appears 
reasonable to expect that over a length scale of order $1/T_c$ (see 
the next section) they should be small compared to unity. On the other 
hand, the value $\sigma=0.04$ is still sufficiently large to allow for 
a relatively rapid onset of the domain formation process, as will 
be seen in section~\ref{sec_num_res}.

\subsection{Coarse graining}\label{subsec_coarse}

A physical length scale is introduced through coarse graining of the 
initial field configuration. For example, a simple algorithm amounts to
replacing the field $\ell(\vec{x})$ at a given site by an average over
a subvolume (box) of size $N_{cg}^3$:
\begin{equation}\label{lcg}
   \ell(\vec{x})\rightarrow \ell'(\vec{x}) =
   \frac{1}{N_{cg}^3}\sum_{\vec{x}' \in {\rm box}} \ell(\vec{x'})\ .
\end{equation}
In physical units, the correlation length at $T_c$ should be on the
order of $1/T_c$ (away from the extreme weak-coupling limit), and 
we set the lattice spacing $a$ by $a\,N_{cg}=1/T_c$.  Dividing 
eqs.~(\ref{eom_lr},\ref{eom_li}) by $T_c^4$ shows that a different 
scale for the initial correlation length corresponds to rescaling 
$Z_s$.

The coarse-graining procedure does not affect the long wavelength part 
of $\ell(\vec{x})$ but reduces the fluctuations,
\begin{equation}\label{e:artcool}
    \langle \delta \ell'^2\rangle < \sigma^2\ .
\end{equation}
Therefore, the counterterms in (\ref{eom_lr},\ref{eom_li}) no
longer match the state of the system after coarse-graining. One needs
to restore the desired fluctuations by rescaling the initial fields to
\begin{equation}\label{e:rescale}
    \delta \ell''(\vec{x})=\delta \ell'(\vec{x})\, \frac{\sigma}
    {\sqrt{\langle\delta \ell'^2\rangle}}\ ,
\end{equation}
so that $\langle \delta \ell''^2\rangle=\sigma^2$.
These fields are taken as initial configuration which is then
propagated in time by solving the equations of motion.

\subsection{Dynamics after a temperature quench}

To perform a quench the temperature is set to a value $T_f$ in the
deconfined phase. Then we use the leapfrog algorithm \cite{Ly05} to
integrate the Euler-Lagrange equations (\ref{eom_lr}), (\ref{eom_li})
in time with the initial conditions described previously. At time $t$,
the structure function is defined by the Fourier transformation of the
Polyakov loop:
\begin{equation}\label{SFdef}
   F(\vec{k},t)= \frac{a^3}{N_s^3}\left|\, \sum_{\vec{x}}
   e^{-i\,\vec{k}\,\vec{x}}\, \ell(\,\vec{x},t\,) \,\right|^2\,.
\end{equation}
For a fixed value of $\vec{k}$, $F(\vec{k},t)$ is called SF. SFs
are our primary observables. In what follows, we label SFs $F_n(t)$ 
similarly as in \cite{BBV06}:
\begin{eqnarray} \label{SFn}
 \vec{k} = \vec{n}\,2\pi/L_s\,, ~~
 n=1:~\vec{n}&=&(1,0,0),\ \vec{n}^{\,2}=1\,, \\
 n=2:~\vec{n}&=&(1,1,0),\ \vec{n}^{\,2}=2\,, \\
 n=3:~\vec{n}&=&(1,1,1),\ \vec{n}^{\,2}=3\,.
\end{eqnarray}
where $L_s\equiv a\,N_s$ denotes the size of the lattice in physical 
units. Note the relation $|\vec{k}|=2\pi \sqrt{n}/L_s$ for 
$n=1,\,2,\,3$. Measurements for $n=1$ include the permutations
$(0,1,0)$, $(0,0,1)$ and for $n=2$ the permutations $(1,0,1)$,
$(0,1,1)$.

\section{Numerical Results}\label{sec_num_res}

We quench to several temperatures in the deconfining phase $T_f/T_c=
1.50$, 1.75, 2.00, 2.25, 2.50 on lattices with spatial extent $N_s=40$, 
48, 64, 80, 96. Periodic boundary conditions are applied and averaging
is done over ensembles of 200 replica. Our length scale set by 
coarse-graining is $aN_{cg}=1/T_c=0.736$~fm, corresponding to the 
SU(3) phase transition temperature of $T_c=260$~MeV~\cite{SU3Tc}. 
Using $N_{cg}=4$ for the correlation length the lattice spacing
\begin{equation} \label{a}
  a=0.184\ {\rm fm}
\end{equation}
follows. Physical volumes $L_s^3=(a\,N_s)^3$ in our simulations are
$(7.4)^3$, $(8.8)^3$, $(11.8)^3$, $(14.7)^3$ and $(17.7)^3$ fm$^3$. To
reduce finite size effects we take lattices that accommodate at least 
10 correlation lengths $N_{cg}$. When we study different physical
volumes, $N_{cg}$ has to be the same for all lattices and $N_{cg}=4$
is a reasonable value. With, say, $N_{cg}=6$ we would have to work on
larger $N_s=60,\cdots,144$ lattices.

\begin{figure} \begin{center}
\includegraphics[width=\columnwidth]{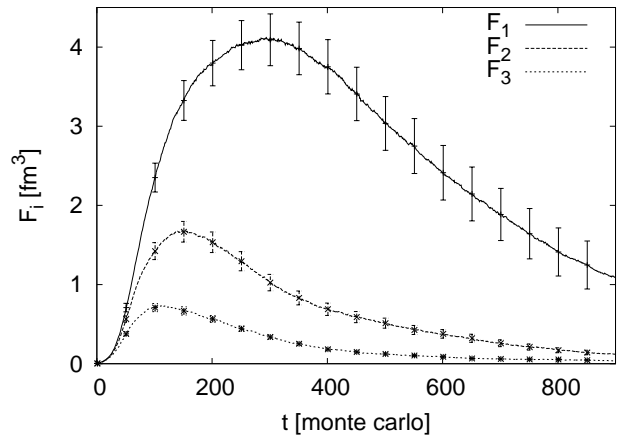}
\caption{Structure factors for Glauber dynamics \cite{BBV06}, $t$ is
  Monte Carlo time in sweeps ($4\times 64^3$ lattice, $T_f/T_c= 1.57$,
  lattice size $L_s=12.1$~fm). $F_n$ corresponds to the $n^{\rm th}$ 
  lattice mode with $|\vec{k}|=2\pi \sqrt{n}/L_s =
  \sqrt{n}\times$102.1~MeV.} \label{fig_sf_glauber}
\end{center} \end{figure}

\begin{figure} \begin{center}
\includegraphics[width=\columnwidth]{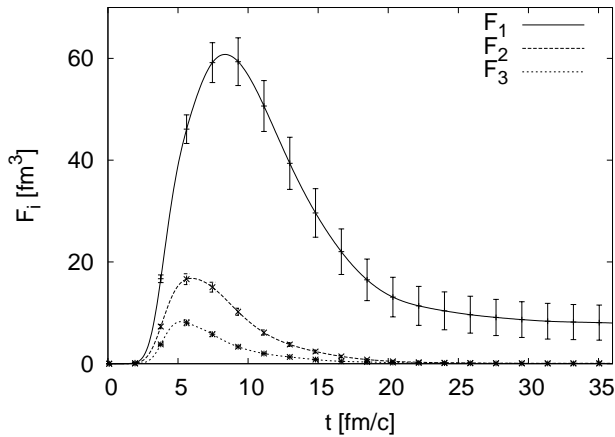}
\caption{Structure factors for Minkowski dynamics, $t$ is real time
($64^3$ lattice, $T_f/T_c=1.5$, lattice size $L_s=11.8$~fm).  $F_n$ 
corresponds to the $n^{\rm th}$ lattice mode with $|\vec{k}|=2\pi 
\sqrt{n}/L_s=\sqrt{n}\times$105.3~MeV.}
\label{fig_sf_mink}
\end{center} \end{figure}

In Fig.~\ref{fig_sf_glauber} we present several SF modes from our
Glauber dynamics study~\cite{BBV06} on a $4\times64^3$ lattice (quench
to $\beta=5.92$, corresponding to $T_f=1.57\,T_c$, average over 170
replica) for comparison with SF modes from the Minkowski dynamics on a
$64^3$ lattice (quench to $T_f=1.50\,T_c$, average over 200 replica)
presented in Fig.~\ref{fig_sf_mink}. The scale on the vertical axis
of Fig.~\ref{fig_sf_glauber} differs from that of
Fig.~\ref{fig_sf_mink} because the former has been determined from the
bare Polyakov loop while the effective Lagrangian~(\ref{ec}) deals
with the renormalized loop. The renormalization constant for the
Polyakov loop~\cite{RenLoop} amounts to a constant multiplying the
SF. Since we are not interested in this renormalization here, this
constant is of minor importance.

\begin{figure} \begin{center}
\includegraphics[width=\columnwidth]{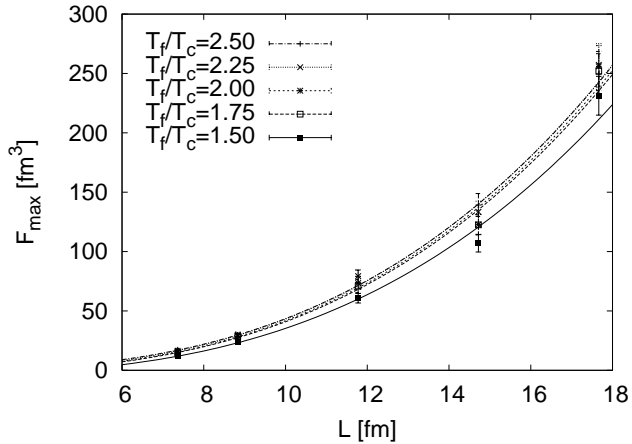}
\caption{Fits of $F_{1,\max}$ for Minkowski dynamics.} 
\label{fig_Fmax} \end{center} 
\end{figure}

Qualitatively, the SFs display the same behavior: an initial
exponential growth is followed by equilibration after the lowest SF
mode reaches its maximum. As for Glauber dynamics, we interpret this
as formation of competing order-order domains between regions of
different Z(3) triality~\cite{Z3domains}, whose equilibration takes a
long time. For the Minkowskian dynamics the maxima of the lowest SF
mode $F_1$ are compiled in Table~\ref{tab_Fmax}. In the normalization
of eq.~(\ref{SFdef}) they scale with volume in the same way as for
Glauber dynamics, provided the volumes are large enough. Fits to the
form $F_{1,\max}=b_0+b_1L_s^3$ are shown in Fig.~\ref{fig_Fmax}.
Interestingly, both Figs.~\ref{fig_sf_glauber} and \ref{fig_sf_mink}
show that the above-mentioned effects due to non-perturbatively large
variations of the Z(3) phase in space are visible even for modes with
$k\sim T/2$. This may be related to Z(3) domain walls forming just 
after the quench being quite broad (disordered phase) and domains 
not much larger than $1/T$.

\begin{table} \centering
\begin{tabular}{|c|c|c|c|c|c|} \hline    
          &$(7.4)^3$&$(8.8)^3$&$(11.8)^3$&$(14.7)^3$&$(17.7)^3$ \\ \hline
$1.50~T_c$ & 12(1) & 24(2) & 60(4) & 107(7) & 231(16)   \\ \hline
$1.75~T_c$ & 15(1) & 29(2) & 71(6) & 122(8) & 252(16)   \\ \hline
$2.00~T_c$ & 15(1) & 30(2) & 74(5) & 123(8) & 256(17)   \\ \hline
$2.25~T_c$ & 16(1) & 29(2) & 79(5) & 133(9) & 257(18)   \\ \hline
$2.50~T_c$ & 17(1) & 28(2) & 79(5) & 139(10)& 249(17)   \\ \hline
\end{tabular}
\caption{$F_{1,\max}$ for different volumes and temperatures
  (Minkowski dynamics).}
\label{tab_Fmax} \end{table}

Since the kinetic term in the Lagrangian (\ref{ec}) is assumed to be
Lorentz invariant, units for time are the same as for length (apart
from the speed of light factor~$c$). When integrating hyperbolic
equations, one uses time steps smaller than the spatial lattice
spacing $a$. We chose $\Delta t/a=0.01$ and, therefore, in physical
units
\begin{equation} \label{Delta_t}
  \Delta t=0.00184\ {\rm fm/c}\ .
\end{equation}
We ran trajectories from 15000 to 25000 time steps corresponding to a
range from 27.6~fm/c to 46~fm/c.

For the case shown in Fig.~\ref{fig_sf_mink} the structure factor for
the first mode takes about 8~fm/c to reach its maximum, and another
$\approx20$~fm/c until that mode equilibrates. On the other hand, the
second and third modes grow for a shorter period of time and
subsequently equilibrate more rapidly. Note that there is an initial
lag of $\approx 2.5$~fm/c, where growth of SFs is only visible on a
logarithmic scale; the precise time for the onset of growth may be 
sensitive to the spectrum and magnitude of initial fluctuations.

\begin{table} \centering
\begin{tabular}{|c|c|c|c|c|c|} \hline
           &$(7.4)^3$&$(8.8)^3$&$(11.8)^3$&$(14.7)^3$&$(17.7)^3$ \\ \hline
$1.50~T_c$ & 5.0(1.2)& 6.3(1.9)& 8.2(1.3) & 11.1(1.4)& 14.0(1.1) \\ \hline
$1.75~T_c$ & 5.4(1.1)& 6.4(1.9)& 9.5(1.4) & 12.0(1.3)& 15.2(1.2) \\ \hline
$2.00~T_c$ & 5.0(1.8)& 7.0(1.5)& 10.1(1.3)& 13.2(1.2)& 16.4(1.0) \\ \hline
$2.25~T_c$ & 5.7(1.5)& 7.4(1.5)& 10.9(1.3)& 14.4(1.1)& 19.1(0.9) \\ \hline
$2.50~T_c$ & 6.1(1.5)& 7.9(1.4)& 11.9(1.3)& 16.2(1.1)& 19.1(0.9) \\ \hline
\end{tabular}
\caption{Times $t_{\max}$ [fm/c] where $F_1(t)$ peaks, for different volumes 
and temperatures (Minkowski dynamics).} \label{tab_tmax} 
\end{table}
In Table~\ref{tab_tmax} we compile the times $t_{\max}$ needed to reach 
the maxima of $F_1$ for different volumes and quench temperatures. 
Evidently, the times before the lowest modes of the system equilibrate 
can be quite large, in $c=1$ units on the order of the size of the 
system. Note also that $t_{\max}$ increases with $T_f/T_c$ because the 
barriers between order-order domains grow higher and are more difficult 
to overcome. This is expected to change for full QCD with light quarks
where the Z(3) symmetry is broken explicitly.

\begin{figure} \begin{center}
\includegraphics[width=\columnwidth]{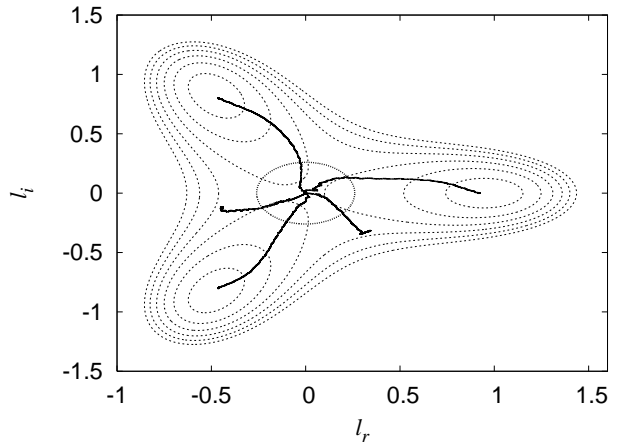}
\caption{Individual trajectories after a quench to $T_f=2.0\,T_c$ on 
a $64^3$ lattice (Minkowski dynamics). Dashed lines show equipotential 
levels from Fig.~\ref{fig_ploop_pot}. The circle about the origin 
indicates the average $t_{\max}$ time.} \label{fig_traj}
\end{center} \end{figure}

\begin{figure} \begin{center}
\includegraphics[width=\columnwidth]{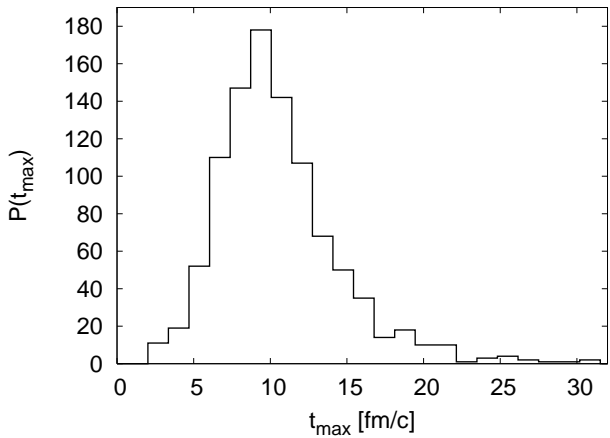}
\caption{Distribution of $t_{\max}$ after a quench to $T_f=2.0\,T_c$ 
on a $64^3$ lattice for an ensemble of 1000 replica (Minkowski dynamics).}
\label{fig_Ptmax} \end{center} \end{figure}

Typical trajectories of the volume-averaged Polyakov loop are shown in
Fig.~\ref{fig_traj}, together with contours of the potential from
Fig.~\ref{fig_ploop_pot}. Some of the configurations in the ensemble
relax directly towards one of the Z(3) minima, while others get stuck
in-between for rather long times. The latter situation corresponds to
the case when there are competing domains of approximately equal size:
the relaxation of the long-wavelength modes is delayed when the
Polyakov loop exhibits non-perturbatively large variations in space.
The circle of radius $c\,t_{\max}$ about the origin indicates the 
average time $t_{\max}$. Note that for some trajectories the 
volume-averaged Polyakov loop (the total ``magnetization'') can 
at $t_{\max}$ still be far from one of the minima of the potential.
Individual configurations exhibit rather large fluctuations about the 
mean, as shown in Fig.~\ref{fig_Ptmax}.

\begin{figure} \begin{center}
\includegraphics[width=\columnwidth]{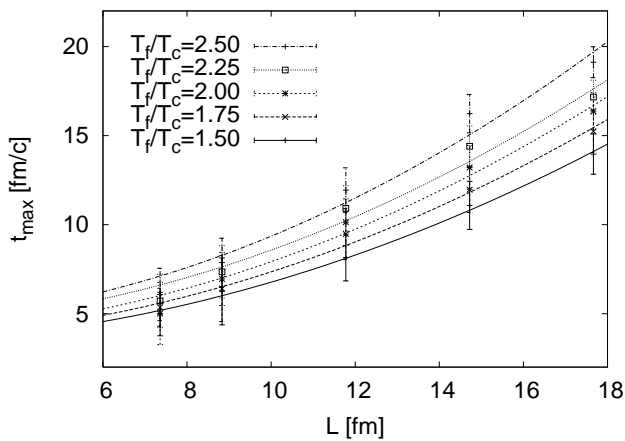}
\caption{Fits of $t_{\max}$ for Minkowski dynamics.} 
\label{fig_tmax} \end{center} \end{figure}
For Glauber dynamics it is known \cite{CL97} that for sufficiently
large systems and corresponding wavelengths $t_{\max}$ scales as
\begin{equation}\label{tmax}
  t_{\max}=a_0+a_1\,L_s^2\,.
\end{equation}
For Minkowski dynamics fits to this form are satisfactory, too, as
shown in Fig.~\ref{fig_tmax}. Thus, in the long wavelength limit the
equilibration time for a mode with wave vector $k$ is proportional to
$1/k^2$. The behavior~(\ref{tmax}) could be used to scale $t_{\max}$
for $F_1$ from table~\ref{tab_tmax} into the corresponding $t_{\max}$
for higher modes via the relation $k_n=2\pi \sqrt{n}/L_s$. For $T_f/T_c
=2.0$, for example, we obtain $a_1=0.044$~fm$^{-1}$ (and $a_0=2.6$~fm, 
which corresponds to the ``lag'' mentioned above). However, given the
rather large statistical error bars shown in Fig.~\ref{fig_tmax} we
cannot exclude a linear dependence,
\begin{equation}
  t_{\max}=a_0+a_1\,L_s~,
\end{equation}
with or without an initial ``lag'' $a_0$, either.

\section{Conclusions and Outlook}\label{sec_concl}

The particle spectrum of a quantum field theory at finite temperature
builds upon the phase of its vacuum ensemble, and the structure of the
vacuum is fundamental for the dynamics of relaxation processes to
equilibrium. In SU(N) gauge theory, non-perturbatively large variations 
of the Z(N) phase within domain walls arise
during the conversion from a confined to a deconfined vacuum
ensemble. We find that the Minkowskian dynamics of a simple model for
SU(3) Polyakov loops (which incorporates the Z(3) structure
of the deconfined vacuum) reproduces the qualitative features of
a previous study \cite{BBV06} of Glauber dynamics within LGT
reasonably well. Heating above $T_c$ drives the
SU(3) gauge theory system from the disordered into the ordered
phase. The initial process of spinodal decomposition is signaled by an
exponential growth of the SFs. Ordering of the system proceeds through
formation of domains of different Z(3) triality until one of them
eventually occupies the whole system (see Fig.~\ref{fig_traj}).

For realistic, sub-asymptotic temperatures as relevant for high-energy
heavy-ion collisions we find that the rather slow dynamics of
competing domains (of the Polyakov loop) delays thermalization of
modes with $k$ nearly up to $T$. In future, it would be interesting 
to quantify the contribution of these modes to the bulk viscosity. 

Our model for Minkowskian dynamics allows us to estimate a physical
time scale for the vacuum conversion process. The relaxation times
found in this way for SU(3) pure gauge theory increase as $\sim1/k^2$
(for sufficiently low $k$) and are estimated to be on the order of the
size $L_s$ of the system for $k=2\pi/L_s$ and $L_s\approx10$~fm. The
hydrodynamic equations describing the evolution of long-wavelength
perturbations in a deconfined medium should therefore be extended to
account for the dynamics of competing domains of the Polyakov loop.

The model for the dynamics of Polyakov loops could be improved in many
ways. As an example, (\ref{ellpot}) deals only with the trace of the
Polyakov loop in the fundamental representation and neglects magnetic
fields.  Effective Lagrangians which include all degrees of freedom of
SU(N) loops as well as magnetic fields (in the static limit) have been
proposed recently~\cite{3deff}. Also, the kinetic term from~(\ref{ec})
for non-equilibrium configurations, for which we have assumed a
Lorentz-invariant form, is in fact not known. Further, one may worry
about the importance of quantum corrections to the dynamics of domain 
walls. Finally, it would be interesting to include the effects due to 
dynamical quarks, which break the center symmetry explicitly. They 
reduce the SF maxima and decrease the relaxation time, as has been 
demonstrated qualitatively for the 3D 3-state Potts model~\cite{BMV04} 
in an external magnetic field, while counter effects may come from the 
accompanying decrease of the transition temperature $T_c$ and the
transformation of the phase transition into a crossover~\cite{QCDTc}.

To improve the simulations, one could drop the assumption of
instantaneous quenching and replace it by the thermalization time of 
the hard modes. The spinodal decomposition process starts already 
during the quench when the central rapidity volume heats up, and it 
occurs in an expanding medium where the longitudinal wave vectors
experience a red-shift~\cite{BottomUp}. Moreover, boundary conditions
in heavy-ion collisions are not periodic as used in this paper;
rather, the surrounding vacuum is confined and therefore disordered.
This increases the width and the effective temperature for the 
deconfinement transition in a volume-dependent way. In a SU(3) 
equilibrium study the increase in temperature was found to be in 
the range from 20\% for a volume of $(5\,{\rm fm})^3$ to 5\% for 
a volume of $(10\,{\rm fm})^3$ \cite{BaBe07}. 

Due to all of the above caveats, our numbers should be viewed only as
rough first estimates. Nevertheless, it appears that away from the
extreme weak-coupling limit ($Z_s\gg1$) the dynamics of competing
domains will influence thermalization of long wavelength modes,
perhaps even for $k$ not very far below $T$, and hence cannot be
neglected. \hfil\break

\centerline{\bf Acknowledgements}

We thank Rob Pisarski for fruitful discussions, and
Brookhaven National Laboratory for its hospitality. This work was in
part supported by DOE grants DE-FG02-97ER-41022, DE-FC02-06ER-41439 and
NSF grant 0555397.


\begin{thebibliography}{99}

\bibitem{QGP} 
I.~Arsene {\it et al.}  [BRAHMS Collaboration],
Nucl.\ Phys.\  A {\bf 757} (2005) 1;
B.~B.~Back {\it et al.},
Nucl.\ Phys.\  A {\bf 757} (2005) 28;
J.~Adams {\it et al.}  [STAR Collaboration],
Nucl.\ Phys.\  A {\bf 757} (2005) 102;
K.~Adcox {\it et al.}  [PHENIX Collaboration],
Nucl.\ Phys.\  A {\bf 757} (2005) 184.

\bibitem{Tc81} L.D. McLerran and B. Svetitsky, Phys. Lett. B 98 (1981) 
          195; J. Kuti, J. Polonyi and K. Szlachanyi, Phys. Lett. B 98
          (1981) 199; J.~Engels, F.~Karsch, I.~Montvay and H.~Satz, 
          Nucl. Phys. B 205 (1982) 545.

\bibitem{SU3Tc} G. Boyd, J. Engels, F. Karsch, E. Laermann, C. Legeland,
                M. L\"utgemeier, and B. Petersson, Nucl. Phys. B 469
                (1996) 419, and references therein.

\bibitem{QCDTc} Z. Fodor, Proc. Science, 42 (Lattice 2007) 011; F. 
                Karsch, Proc. Science, 42 (Lattice 2007) 015.

\bibitem{MG_McL} See, for example, M.~Gyulassy and L.~McLerran,
                 Nucl.\ Phys.\ A {\bf 750} (2005) 30.

\bibitem{BottomUp}
R.~Baier, A.~H.~Mueller, D.~Schiff and D.~T.~Son,
Phys.\ Lett.\  B {\bf 502} (2001) 51.

\bibitem{ALM04}
P.~Arnold, J.~Lenaghan and G.~D.~Moore,
JHEP {\bf 0308} (2003) 002.

\bibitem{tHooft}
G.~'t Hooft,
Nucl.\ Phys.\  B {\bf 138} (1978) 1;
Nucl.\ Phys.\  B {\bf 153} (1979) 141;
A.~M.~Polyakov,
Phys.\ Lett.\  B {\bf 72} (1978) 477.

\bibitem{Z3domains}
T.~Bhattacharya, A.~Gocksch, C.~Korthals Altes and R.~D.~Pisarski,
Nucl.\ Phys.\  B {\bf 383} (1992) 497;
C.~P.~Korthals Altes,
Nucl.\ Phys.\  B {\bf 420} (1994) 637;
C.~Korthals-Altes, A.~Kovner and M.~A.~Stephanov,
Phys.\ Lett.\  B {\bf 469} (1999) 205;
P.~Giovannangeli and C.~P.~Korthals Altes,
Nucl.\ Phys.\  B {\bf 608} (2001) 203;
Nucl.\ Phys.\  B {\bf 721} (2005) 1;
Nucl.\ Phys.\  B {\bf 721} (2005) 25;
P.~de Forcrand, M.~D'Elia and M.~Pepe,
Phys.\ Rev.\ Lett.\  {\bf 86} (2001) 1438;
P.~de Forcrand and L.~von Smekal,
Phys.\ Rev.\  D {\bf 66} (2002) 011504;
P.~de Forcrand and D.~Noth,
Phys.\ Rev.\  D {\bf 72} (2005) 114501;
P.~de Forcrand, C.~Korthals-Altes and O.~Philipsen,
Nucl.\ Phys.\  B {\bf 742} (2006) 124.

\bibitem{BBV06} A. Bazavov, B.A. Berg, and A. Velytsky, Phys. Rev.
                D 74 (2006) 014501.

\bibitem{MO00} T.R. Miller and M.C. Ogilvie, Phys. Lett. B 488 (2000) 
               313; Nucl. Phys. B (Proc. Suppl.) 94 (2001) 419. As in
               these references we use the terminology ``spinodal
               decomposition'' in a broader sense than some 
               statistical physicists do.

\bibitem{Gl63} R.J. Glauber, J. Math. Phys. 4 (1963) 294. Model~A in
               the classification of P.M. Chaikin and T.C. Lubensky,
               {\it Principles of condensed matter physics}, Cambridge
               University Press, Cambridge 1997, Table~8.61.1, p.467.

\bibitem{BBV06a} A. Bazavov, B.A. Berg, and A. Velytsky, Proc. Science
                 32 (Lattice 2006) 127.

\bibitem{SY82} B. Svetitsky and L.G. Yaffe, Nucl. Phys. B 210 (1982) 423.

\bibitem{CL97} P.M. Chaikin and T.C Lubensky, p.484 (ref.~\cite{Gl63}).

\bibitem{Pi00} 
R.D. Pisarski, Phys. Rev. D 62 (2000) 111501(R); 
A.~Dumitru and R.~D.~Pisarski,
Phys.\ Lett.\  B {\bf 504}, 282 (2001);
Phys.\ Rev.\  D {\bf 66} (2002) 096003.

\bibitem{MFK07} A.J. Mizher, E.S. Fraga, G. Krein, Braz. J. Phys. 37 
                (2007) 605; N.C. Cassol-Seewald, R.L.S. Farias, E.S. 
                Fraga, G. Krein, and R.O. Ramos, arXiv:0711.1866v1.

\bibitem{Du01} O. Scavenius, A. Dumitru, and A.D. Jackson, Phys. Rev.
               Lett. 87 (2001) 182302.

\bibitem{Langevin}
E.~S.~Fraga, G.~Krein and A.~J.~Mizher,
Phys.\ Rev.\  D {\bf 76} (2007) 034501.

\bibitem{Kpi} O.~Scavenius, A.~Dumitru and J.~T.~Lenaghan, Phys.\ Rev.\  
              C {\bf 66} (2002) 034903. 

\bibitem{TV05} E.T. Tomboulis and A. Velytsky, Phys. Rev. D 72 (2005)
               074509 and references therein.

\bibitem{Ly05} D. Frenkel and B. Smit, {\it Understanding Molecular
               Simulation}, Academic Press, 2002.

\bibitem{RenLoop} O.~Kaczmarek, F.~Karsch, P.~Petreczky and F.~Zantow,
Phys.\ Lett.\  B {\bf 543} (2002) 41;
A.~Dumitru, Y.~Hatta, J.~Lenaghan, K.~Orginos and R.~D.~Pisarski,
Phys.\ Rev.\  D {\bf 70}  (2004) 034511;
S.~Gupta, K.~H\"ubner and O.~Kaczmarek,
Phys.\ Rev.\  D {\bf 77} (2008) 034503.

\bibitem{3deff}
A.~Vuorinen and L.~G.~Yaffe,
Phys.\ Rev.\  D {\bf 74} (2006) 025011;
R.~D.~Pisarski,
Phys.\ Rev.\  D {\bf 74} (2006) 121703;
Ph.~de Forcrand, A.~Kurkela and A.~Vuorinen,
arXiv:0801.1566 [hep-ph].

\bibitem{BMV04} B.A. Berg, H. Meyer-Ortmanns, and A. Velytsky, Phys. Rev.
                D 70 (2004) 054505.

\bibitem{BaBe07} A. Bazavov and B.A. Berg, Phys. Rev. D 76 (2007) 014502.

\end{thebibliography}
\end{document}